\documentstyle[12pt]{article}

\begin{document}
\begin{center}
{\bf Antiferromagnetic Spin Ladders with \\Odd and Even number of Chains}
\vskip 1cm
Asimkumar Ghosh and Indrani Bose\\
Department of Physics,\\ Bose Institute,\\
93/1, Acharya Prafulla Chandra Road,\\ Calcutta-700 009, India.
\end{center}
\vskip .8cm
\begin{abstract}
 We construct frustrated antiferromagnetic spin ladders with m chains for 
which the exact ground state can be determined in a particular parameter 
regime. The excitation spectrum is shown rigorously to be gapless ( with gap ) 
for odd ( even ) m. In a general parameter regime, the four-chain and 
periodic ladders are studied using a mean-field theory based on the bond 
operator formalism for spin $S\,=\,\frac{1}{2}$. The excitation spectrum 
and the spin-gap are calculated in both the cases. The spin-gap of the 
frustrated ladder system has a larger magnitude than in the case of 
non-frustrated ladders. For the frustrated periodic ladder, the spin-gap 
vanishes at a critical value of the interladder coupling strength 
which is larger than the critical value in the case of 
non-frustrated periodic ladder.
\end{abstract}
\vskip 1cm
P.A.C.S. NO:   75.10.Jm
\section*{I. Introduction}
Antiferromagnetic (AFM) spin S=1/2 ladders have been extensively studied 
in recent times as the ladder interpolates between the 1d chain and 2d 
plane.\cite{Dagotto_Rice} The S=1/2 Heisenberg AFM chain is exactly solvable 
using the well-known Bethe Ansatz.\cite{Bethe} The ground state is disordered 
and characterised by a power law decay of the two-spin correlation function. 
The excitation spectrum is known to be gapless. The planar (square lattice) 
S=1/2 AFM with nearest-neighbour coupling shows long range AFM order 
at zero temperature and the excitation spectrum is gapless.

The copper-oxide planes of the high-$T_c$ cuprate systems in the undoped 
state serve as good examples of 2d AFMs defined on the square lattice. 
This fact has sparked renewed interest in the study of low-dimensional 
AFMs. One particularly interesting problem is to study the crossover from AFM 
chains to the square lattice. The crossover can be understood by examining 
n-chain spin ladders with increasing width. Such studies have yielded 
the surprising result that for odd (even) values of n, the excitation 
spectrum is gapless (with gap). Ladders consisting of an even number 
of chains have spin liquid ground state with exponential decay of 
the two-spin correlation function. The spin 1 excitation of the ladder 
has a finite energy gap. A ladder with an odd number of chains has quite 
different behaviour and displays characteristics similar to those of 
single chains, namely, gapless spin excitations and a power-law decay 
of the two-spin correlations. The significant difference between 
even-chain and odd-chain ladders can be attributed to quantum effects.

The compound $(VO)_2P_2O_7$ has a two-chain ladder configuration of 
spin-1/2  $V^{4+}$ ions.\cite{Johnston} Real compounds like stoichiometric 
$Sr_{n-1}Cu_{n+1}O_{2n}, (n=3,5,7,9,...)$ \cite{Gopalan} can be described by m-chain 
spin ladders with $m\,=\,\frac{n+1}{2}$. Spin susceptibility and nuclear 
magnetic resonance (NMR) experiments on the two-chain ladder systems 
show the existence of a gap in the excitation spectrum. Neutron 
scattering and muon spin resonance experiments give clear signs of 
short-range spin order in the 2-chain ladders.\cite{Dagotto_Rice} Three-chain ladders 
$(Sr_2Cu_3O_5)$ by contrast show longer range spin correlations 
and a gapless spectrum. There is also true long range order at 
low temperatures brought about by weak interladder interactions.\cite{Dagotto_Rice}

Some theoretical studies have been undertaken recently to understand 
the `odd-even' effect of spin ladders. Reigrotzki et al \cite{Reigro..} 
have studied the properties of spin ladders with two, three, and four 
chains expanded in the ratio of the intrachain and interchain coupling 
constants. Khveshchenko\cite{Khves..} has shown that for odd-chain ladders 
a topological term appears in the effective action corresponding to the 
dynamics at long wavelengths. For even-chain ladders there is no such 
term. This topological term is similar to the one responsible for the 
difference between integer and half-odd integer spin chains. Integer 
spin chains have a gap ( the Haldane gap ) in the excitaion spectrum 
whereas half-odd integer spin chains are gapless. Recent studies \cite{Hida,Steven}
have shown that two-chain spin ladders with both ferromagnetic and 
antiferromagnetic rung exchange interactions are Haldane gap systems 
in the appropriate limits. Rojo \cite{Rojo} has further given a rigorous 
proof for the absence of gap for spin 1/2 ladders with an odd number 
of chains in the infinite chain length limit.

In this paper, we construct models of spin ladders with odd and even 
number of chains for which the exact ground states can be determined 
and the `odd-even' effect associated with the excitation spectrum of 
ladders can be demonstrated rigorously. In Sec. II,
a description of the ladders is given and the ground and first excited 
state determined. In Sec. III,
the properties of a four-chain ladder are studied using a mean field 
theory based on the bond operator formalism\cite{Gopalan,Sachdev}. 
The calculations are further extended to periodic ladders. Sec. IV
contains concluding remarks.

\section*{II. Model spin ladders}
Bose and Gayen \cite{Bose_Gayen1,Bose_Gayen2,Gayen_Bose1} have constructed a two-chain spin ladder 
model for which several exact results can be derived both in the undoped 
and hole-doped states. The ladder is shown in Fig. 1. Every site is 
occupied by a spin of magnitude 1/2. The spins interact with Heisenberg 
AFM exchange interaction. The nearest-neighbour (n.n.) intra-chain exchange 
interaction is of strength $\beta$, the rung and n.n. diagonal exchange 
interactions are of strength $\alpha$ and $\gamma$ respectively. For $\beta 
= \gamma$ and $\frac{\alpha}{\beta}\geq 2$, Bose and Gayen\cite{Bose_Gayen1} showed that 
the exact ground state consists of singlet spin configurations $\left(\frac{1}{\sqrt{2}}
(\uparrow \downarrow - \downarrow \uparrow)\right)$ along the rungs of the ladder. We 
construct spin ladders of increasing width by adding chains and stipulating 
that alternate two-chain ladders have no diagonal exchange inteactions. The 
spins located in the rungs of the model interact via the `sawtooth-chain' 
interaction. Such a chain is illustrated in Fig. 2. In Figs. 3(a) and 3(b), 
the spin ladders with three and four chains respectively are shown. For both 
odd and even-chain spin ladders one can write down an exact eigenstate which 
is also the exact ground state in an appropriate parameter regime. The exact 
eigenstate for an even-chain spin ladder consists of spin singlets along 
the rungs of two-chain ladders with diagonal exchange interactions. The proof 
of eigenstate can be easily obtained using the spin identity $\vec{S}_n\cdot
(\vec{S}_l\, + \,\vec{S}_m)\,\big[lm\big]\equiv 0$ where \big[lm\big] describes a spin singlet of 
spins at sites l and m. Let \(\eta\) be the strength of the exchange interaction 
between the next-nearest-neighbour (n.n.n.) spins along the rungs. The rung 
exchange interactions for two chain ladders without diagonal interactions 
are of strength \(\xi\). Consider the parameter regime \(\beta = \gamma = \xi = \eta\)
and $\frac{\alpha}{\beta}\, \geq \,3$. In this regime, the exact eigenstate 
described before becomes the exact ground state. The proof of the exact 
ground state can be obtained by using the well-known method of `divide 
and conquer'.\cite{Bose_Gayen1} Let $E_1$ be energy of the exact eigenstate and 
$E_G$ the energy of the exact ground state. Then $E_G\, \leq \,E_1$. For the 
specified eigenstate  $E_1\, =\, \frac{-3N\alpha}{4}$  where N is the total 
number of n.n. vertical bonds (rungs) in the system along which singlets 
form in the ground state. The total spin 
Hamiltonian H can be divided into two parts $H_L$ and $H_S$. $H_L$ 
describes the exchange interactions of all the two-chain ladders in the system 
with diagonal interactions. Each such ladder has exchange interaction 
strengths $\alpha_1\,(\,\alpha_1\,\geq\,2\beta\,)$, $\beta$ and $\beta$ for the rung, intra-chain n.n. 
and diagonal interactions respectively. The exact ground state energy 
$E_L\,=\,\frac{ -3N\alpha_1}{4}$. $H_S$ corresponds to all the rung interactions 
in the system having the geometrical structure of sawtooth chains. All the spin-spin 
interactions in the sawtooth are of strength $\beta$. The ground state energy 
$E_S \,=\, \frac{-3N\beta}{4}$. Let $\Psi_G$ be the exact ground state wave function. 
Then by variational theorem,
\begin{eqnarray*}
\langle\Psi_G|H|\Psi_G\rangle \,=\, E_G \,=\, \{\langle \Psi_G|H_L|\Psi_G\rangle
\,+\,\langle\Psi_G|H_S|\Psi_G\rangle\}
\end{eqnarray*}
i.e.,  $E_1 \,\geq\, E_G \,\geq\, E_L\, +\, E_S$.
\vskip .15cm
Now, $E_1 \,= \,\frac{-3N \,\alpha}{4} \,(\,\alpha \,=\, \alpha_1\,+\,\beta\,\geq\,3\,\beta\,) \,=\, 
E_L \,+\, E_S$. So $E_1\,=\, E_G$ 
and the exact eigenstate is also the exact ground state. For a spin ladder 
of odd number of chains, all the chains except one belong to the two-chain 
ladders with diagonal exchange interactions. Again, in the parameter regime 
$\alpha \,\geq \,3 \beta$ and $\beta \,=\, \gamma \,=\, \xi \,=\, \eta$, the exact ground state 
consists of spin singlets along the rungs of the two-chain ladders with 
diagonal interactions and the isolated chain has spin configuration 
corresponding to that of the Heisenberg AFM chain. The proof of exact ground 
state is similar to that for even-chain ladders.

We now consider the excited states of the system. For odd-chain ladders, the 
lowest lying excited state is the triplet \((S = 1)\) excitation of the HAFM 
chain. The excitation energy is given by \cite{Cloizeaux}
\begin{eqnarray*}
\omega =\frac{ \pi}{2}\,\beta \,|sin(q)| \quad \quad \quad \quad \quad \quad (2.1)
\end{eqnarray*}
where q is the momentum wave vector w.r.t that of the chain ground state. 
The excitation is confined to the chain which does not belong to the 
two-chain ladders with diagonal couplings. The spectrum is gapless for $q=0$ and $\pi$. 
In the spin-ladder systems considered, 
periodic boundary condition is assumed to hold true in the horizontal (x) direction. 
The ladder has infinite length  in this direction. For the two-chain spin ladder 
shown in Fig. 1 and in the parameter regime under consideration, the lowest 
excited states consist of a triplet along one of the rungs. The excitation energy 
measured w.r.t the ground state energy is $\alpha$ which is a measure of the 
spin-gap. The triplet excitation is localised and has no dynamics in the 
x-direction. For an even-chain ladder, the spin dynamics is only in the vertical 
y-direction and the lowest excited state corresponds to that of the sawtooth chain. 
Consider the four-chain ladder shown in Fig. 3(b). Exact diagonalisation of 
four-sited sawtooth chain shows that the first excited state has energy 
$-\,\frac{(1 \,+\, \sqrt{33})}{4}\beta$. The energy measured w.r.t the ground state 
energy, $-\,\frac{18\,\beta}{4}$, gives a spin-gap which is less than $\alpha = 3\beta$, 
the spin gap for the two-chain ladder. The spin-gap thus decreases in 
magnitude as the number of chains in the even-chain ladder increases from 
two to four. In the next section, we consider more general parameter regimes 
in which the exact ground and excited states are not known. We study the 
even-chain ladders only and determine the excitation spectrum and spin-gap 
for both the four-chain and periodic ladders.
\section*{III. Four-chain and periodic ladder}
The properties of a two-chain spin ladder are already known in the mean-field 
theory.\cite{Gayen_Bose2} The Hamiltonian is given by (Fig.1)
\begin{eqnarray*}
H\,&=&\, \sum_{i}\, \{ \alpha \,{\bf S}_i \cdot {\bf S}_i^{\prime} + \beta\,({\bf S}_i^{\prime} \cdot {\bf S}_{i+1}^{\prime} +
{\bf S}_i \cdot {\bf S}_{i+1})\\ \\
\,&+&\,\gamma\,({\bf S}_i \cdot {\bf S}_{i+1}^{\prime} + {\bf S}_i^{\prime} \cdot {\bf S}_{i+1})\}
\quad\quad\quad\quad\quad(3.1)
\end{eqnarray*}
The ground state is assumed to be in a dimerized phase with the singlet 
dimers located along the rungs. The bond operator representation of  {\bf S}=1/2 
spins is used to study the properties of dimerized phases. We consider two spins 
(S=1/2) $ S_i^\prime$ and $S_i$ placed on each rung. The Hilbert space 
consists of four states which in appropriate combinations describe the singlet 
$\bf |s\rangle$ and the three triplet $\bf |t_x\rangle, \,|t_y\rangle \,$ and $\bf  |t_z\rangle$ 
states. These states are created out of the vacuum $|0\rangle$ by the singlet 
and triplet creation operators
\begin{eqnarray*}
|{\bf s} \rangle \,=\, {\bf s}^{\dag} |0 \rangle &=& \frac{1}{\sqrt{2}}(| \uparrow \downarrow \rangle -
| \downarrow \uparrow \rangle) \\ \\
|{\bf t}_x \rangle = {\bf t}_x^{\dag} |0 \rangle &=& -\,\,\frac{1}{\sqrt{2}}(| \uparrow \uparrow \rangle - 
| \downarrow \downarrow \rangle) \\ \\
|{\bf t}_y \rangle = {\bf t}_y^{\dag} |0 \rangle &=& \frac{i}{\sqrt{2}}(| \uparrow \uparrow \rangle +  
| \downarrow \downarrow \rangle) \\ \\
|{\bf t}_z \rangle = {\bf t}_z^{\dag} |0 \rangle &=& \frac{1}{\sqrt{2}}(| \uparrow \downarrow \rangle + 
| \downarrow \uparrow \rangle)\quad \quad \quad \quad \quad (3.2)
\end{eqnarray*}
The spins $S_i^\prime$ and $S_i$, in terms of the singlet and triplet spin 
operators, are given by \cite{Gopalan,Sachdev}
\begin{eqnarray*}
S_{i \alpha}^{\prime}\,&=&\,\frac{1}{2} \,({\bf s_i^{\dag}}\, {\bf t}_{i \alpha}\,+\,{\bf t}_{i \alpha}^{\dag} \, {\bf s_{i}} -
i {\bf \epsilon}_{\alpha \beta \gamma} \, {\bf t}_{i \beta}^{\dag} \, {\bf t}_{i \gamma}) \quad \quad \quad (3.3)\\ \\
S_{i \alpha}\,&=&\,\frac{1}{2}({\bf -s_i^{\dag}} \, {\bf t}_{i \alpha} - {\bf t}_{i \alpha}^{\dag} \, {\bf s_{i}} -
i{\bf \epsilon}_{\alpha \beta \gamma}\,{\bf t}_{i \beta}^{\dag} \,{\bf t}_{i \gamma})\quad \quad \quad (3.4)
\end{eqnarray*}
${\bf \alpha, \,\beta \,}$ and ${\bf \gamma}$ are the components along the x, y and z axes 
respectively and ${\bf \epsilon}$ is the Levi-Civit\`{a} symbol and represents the 
totally antisymmetric tensor. All repeated indices over ${\bf \alpha, \,\beta \,}$ and ${\bf \gamma}$ 
are assumed to be summed over.

A constraint of the form
\begin{eqnarray*}
{\bf s}^{\dag} {\bf s} \,+\, {\bf t}_{\alpha}^{\dag} {\bf t}_{\alpha}\,=\,1\quad \quad\quad\quad\quad \quad \quad \quad \quad(3.5)
\end{eqnarray*}
is assumed to hold true for each dimer so that the physical states can be 
either singlets or triplets. The singlet and triplet operators at each site satisfy bosonic commutation 
relations
\begin{eqnarray*}
[{\bf s,s}^{\dag}]\,=\,1,\quad [{\bf t_{\alpha},t_{\beta}^{\dag}}]\,=\,\delta_{\alpha \beta},\quad [{\bf s,t_{\alpha}^{\dag}}]\,=\,0 
\quad \quad \quad \quad (3.6)
\end{eqnarray*}
One now substitutes the operator representation of spins given in Eqs.(3.3) and 
(3.4) into the original Hamiltonian (Eq.(3.1)). 
A site-dependent chemical potential ${\bf \mu_i}$ is included in the 
Hamiltonian to impose the constraint of Eq.(3.5). The transformed Hamiltonian 
can be solved by a mean-field decoupling of the quartic terms containing two {\bf s} 
and two {\bf t} operators as well as four {\bf t} operators. One takes $\langle {\bf s}_i \rangle \,=\,\bar{{\bf s}}$ and replaces the 
local constraint $\bf \mu_i$ by a global one $\bf \mu$. One also defines two mean fields as
\begin{eqnarray*}
P \,=\,\langle {\bf t_{i\alpha}^{\dag} \, t_{i+1, \alpha}} \,\rangle \quad
Q \,=\,\langle {\bf t_{i \alpha} \, t_{i+1 \alpha}} \rangle \quad\quad\quad\quad\quad (3.7)
\end{eqnarray*}
Next, a Fourier transformation 
of the operators is taken. The resultant Hamiltonian can be diagonalised by 
the Bogolyubov transformation. Since the details of the calculation are available 
elsewhere \cite{Gopalan} we quote the final results. The diagonalised mean-field 
Hamiltonian ${\bf H_m(\mu,\,\bar{s},\,P,\,Q)}$ is given by 
\begin{eqnarray*}
H_m\left({\bf \mu,\,\bar{s},\,P,\,Q}\right)\,&=&\,N\,\left(\,-\frac{3}{4}{\bf \bar{s}^2 \alpha \,-\,\mu} \, \bar{s}^2 \, + \,\mu\right)
\,- \,\frac{N}{2} \, \left(\frac{\alpha}{4} \,- \,{\bf \mu}\right)\\ \\
 \,&-&\,\frac{N\lambda_2}{3}\left(P^2\,-\,Q^2\right) 
\,+ \, \sum_k \omega_k\left( \gamma_k^{\dag}\gamma_k \, + \,\frac{1}{2}\right) \quad \quad (3.8)
\end{eqnarray*}
where
\begin{eqnarray*}
\omega_k \,&=&\,\sqrt{\Lambda_k^2 \,- \,(2\, \Delta_k\,)^2\,} \quad \quad \quad \quad \quad \quad\quad\quad\quad\quad(3.9)\\ \\
\Lambda_k \,&=&\,\left(\frac{\alpha}{4} \,- \,\mu\right) \,+ \,\left(\lambda_{1}\,\,{\bf \bar{s}^2 } \,+\,\frac{2\,P\,\lambda_2}{3}\right)\cos{k} \\ \\
\Delta_k\,&=&\,\left( \frac{\lambda_{1}\,\,{\bf \bar{s}^2}}{2}\,-\,\frac{Q\,\lambda_{2}}{3}\right) \cos{k} \\ \\
\lambda_{1}\,&=&\,\left(\beta \,- \,\gamma\right)\\ \\
\lambda_{2}\,&=&\,\left(\beta\,+\gamma\right)\quad \quad\quad\quad \quad \quad \quad \quad\quad(3.10)
\end{eqnarray*}
The parameters $\bf \mu,\,\bf {\bar{s}},\, P $ and {\bf Q} can be determined from appropriate 
self-consistent equations and the spin-gap $\Delta$ is given by
\begin{eqnarray*}
\Delta \,= \,\sqrt{\left(\frac{\alpha}{4}-\mu-\frac{2\,\lambda_{2}}{3}\left(P\,+\,Q\right)\right)\,
\left(\frac{\alpha}{4}-\mu\,-\,2\,\lambda_{1}\,{\bf {\bar s}}^2\,-\,\frac{2\,\lambda_{2}}{3}\left(P\,-\,Q \right)\right)}
\\   \quad \quad \quad \quad (3.11)
\end{eqnarray*}
We now consider the four-chain spin ladder shown in Fig.3(b). The number 
of two-chain spin ladders is two and they are designated as left (top) and 
right (bottom) ladders. The two ladders are coupled by exchange interaction 
of strength $\bf {\xi}$. The hamiltonian describing the system is given by
\begin{eqnarray*}
H\,&=&\,\sum_i \{ \alpha ({\bf S}_{l\,i} \cdot {\bf S}_{l\,i}^{\prime}\,
+\,{\bf S}_{r\,i} \cdot {\bf S}_{r\,i}^{\prime} )+
\beta ({\bf S}_{l\,i}^{\prime} \cdot {\bf S}_{l\,i+1}^{\prime} \,
+\,{\bf S}_{l\,i} \cdot {\bf S}_{l\,i+1}\\ \\
\,&+&\,
{\bf S}_{r\,i}^{\prime} \cdot {\bf S}_{r\,i+1}^{\prime}\,+\,{\bf S}_{r\,i} \cdot {\bf S}_{r\,i+1}) \, 
\,+\,\gamma ({\bf S}_{l\,i}^{\prime} \cdot {\bf S}_{l\,i+1}\,
+\,{\bf S}_{l\,i}\cdot {\bf S}_{l\,i+1}^{\prime}\\ \\
\,&+&\,{\bf S}_{r\,i}^{\prime} \cdot {\bf S}_{r\,i+1}\,+\,{\bf S}_{r\,i}\cdot {\bf S}_{r\,i+1}^{\prime} ) +
\eta \,{\bf S}_{l\,i}^{\prime} \cdot {\bf S}_{r\,i}^{\prime}\,
+\,\xi \,{\bf S}_{l\,i}\cdot {\bf S}_{r\,i}^{\prime} \}  \\ 
&&\quad\quad\quad\quad\quad\quad\quad \quad\quad\quad\quad\quad\quad\quad\quad\quad\quad\quad\quad (3.12)
\end{eqnarray*}
The spin operators are expressed in terms of the singlet and triplet bond 
operators through the transformations given in Eqs.(3.3) and (3.4). The transformed 
Hamiltonian is given by
\begin{eqnarray*}
H\,&=&\,\sum_{i} \{ \sum_{m=l,r}[\alpha \,\left(-\frac{3}{4} 
\,{\bf s}_{m\,i}^{\dag} {\bf s}_{m\,i} \,+ \,\frac{1}{4}\,
{\bf t}_{m \, i \,\alpha}^{\dag}\,{\bf t}_{m \, i \,\alpha} \right)\\ \\
\,&-& \,{\bf \mu}_{m\,i}\left({\bf s}_{m\,i}^{\dag} {\bf s}_{m\,i}\,+\,{\bf t}_{m \,i \,
\alpha}^{\dag} {\bf t}_{m \,i \,\alpha}\,-\,1 \right) \\ \\
\,&+&\,\frac{\lambda_{1}}{2}\left({\bf t}_{m\,i\,\alpha}^{\dag} 
{\bf t}_{m\,i+1\,\alpha}s_{m\,i+1}^{\dag} {\bf s}_{m\,i}
\,+\, {\bf t}_{m\,i\,\alpha}^{\dag} {\bf t}_{m\,i+1\,\alpha}^{\dag} 
{\bf s}_{m\,i+1} {\bf s}_{m\,i} \,+\,H.C.\,\right) \\ \\
\,&-&\,\frac{\lambda_{2}}{2}\,{\bf \epsilon}_{\alpha\,\beta\,\gamma}\,
{\bf \epsilon}_{\alpha\,\beta^{\prime}\,\gamma^{\prime}}\,
{\bf t}_{m\,i\,\beta}^{\dag}
\,{\bf t}_{m\,i+1\,\gamma}\,{\bf t}_{m\,i\,\beta^{\prime}}^{\dag}\,
{\bf t}_{m\,i+1\,\gamma^{\prime}} ] \\ \\
\,&+&\,\frac{\lambda_{3}}{4}\left({\bf s}_{l\,i}^{\dag} \,
{\bf t}_{l\,i\,\alpha}\, {\bf t}_{r\,i\,\alpha}\,{\bf s}_{r\,i}^{\dag}\,+\,
{\bf s}_{l\,i}^{\dag} \,{\bf t}_{l\,i\,\alpha}\,{\bf t}_{r\,i\,\alpha}^{\dag}\,{\bf s}_{r\,i}\,
+\,H.C.\right) \\ \\
\,&-&\,\frac{\lambda_{4}}{4}\,{\bf \epsilon}_{\alpha\,\beta\,\gamma}\,
{\bf \epsilon}_{\alpha\,\beta^{\prime}\,\gamma^{\prime}}\,
{\bf t}_{l\,i\,\beta}^{\dag}\,{\bf t}_{l\,i\,\gamma}\,
{\bf t}_{r\,i\,\beta^{\prime}}^{\dag}\,{\bf t}_{r\,i\,\gamma^{\prime}} \} 
\quad\quad\quad\quad\quad\quad\quad \quad \quad \quad (3.13)
\end{eqnarray*}
where
\begin{eqnarray*}
\lambda_1\,=\,\beta\,-\gamma, \quad\lambda_2\,=\,\beta+\gamma,
\quad\lambda_3\,=\,\eta\,-\,\xi, \quad \lambda_4\,=\,\eta\,+\,\xi \quad\quad(3.14)
\end{eqnarray*}
where m denotes the ladder index, left (l) or right (r) and ${\bf \mu_{mi}}$ is the 
chemical potential which has been introduced to take account of the constraint 
specified in Eq.(3.5). 
One takes the expectation value 
$\bf {\langle s_{m i}\rangle\,=\,\bar{s}}$ and the local chemical potential $\bf {\mu_{m i}}$ 
is replaced by the global one $\bf {\mu}$. We perform a Fourier transformation of the 
operators $\bf {t_{m i \alpha}\,=\,\frac{1}{\sqrt{N}}\sum_k t_{m k \alpha}e^{-ikr_i}}$ 
where N is the number of dimers or rungs in a two chain ladder and k is the 
wave vector along the ladder axis. The Fourier-transformed Hamiltonian is 
given by 
\begin{eqnarray*}
H\,\,&=&\,\,2N \,\left(- \,\frac{3}{4} \,{\bf \alpha \, \bar{s}}^2 \, - \,{\bf \mu \, \bar{s}}^2 \,
+ \,{\bf \mu} \right)\\\\
 \,&-&\,\frac{2}{3}\,\lambda_{2}\,N\,(P^2\,-\,Q^2)\,
-\,\frac{1}{6}\,\lambda_{4}\,N\,(P^{\prime\,2}\,-\,Q^{\prime\,2})  \\ \\
\,&+&\,\sum_k \,[\,\sum_{m=l,\,r}\,\left(A_k \,{\bf t}_{m\,k\,\alpha}^{\dag} 
{\bf t}_{m k \alpha}\,+\,B_k\,\left({\bf t}_{mk\alpha}^{\dag}\,{\bf t}_{m-k\alpha}^{\dag} 
\,+\, {\bf t}_{m\,k\,\alpha}\,{\bf t}_{m\,-k\,\alpha}\right)\right) \\ \\
\,&+&\,C\, \left({\bf t}_{l\,k\,\alpha}^{\dag} 
{\bf t}_{r\,k\,\alpha} \,+\, {\bf t}_{r\,k\,\alpha}^{\dag}\,{\bf t}_{l\,k\,\alpha}\right)
\,+\,D\,\left({\bf t}_{l\,k\,\alpha}^{\dag}\,{\bf t}_{r\,-k\,\alpha}^{\dag}\,
+\,{\bf t}_{l\,k\,\alpha} {\bf t}_{r\,- k\,\alpha}\right)\,\,] 
\,\,\, (3.15)
\end{eqnarray*}
where $A_k,\,B_k$, C and D are defined as
\begin{eqnarray*}
A_k\,&=&\,\frac{\alpha}{4}\,-\,\mu\,+\,\left(\lambda_{1}\,{\bf {\bar s}}^2\,
\,+\,\frac{2}{3}\,\lambda_{2}\,P\,\right)\cos{k}\,\,,\quad \\ \\
B_k\,&=&\,\left(\frac{1}{2}\,\lambda_{1}\,{\bf {\bar s}}^2
\,-\,\frac{1}{3}\,\lambda_{2}\,Q\right)\,\cos{k}\,\,,\\ \\
C\,&=&\,\frac{1}{4}\,\lambda_{3}\,{\bf {\bar s}}^2
\,+\,\frac{1}{6}\,\lambda_{4}\,P^{\prime}\,,\\ \\ 
D\,&=&\,\frac{1}{4}\,\lambda_{3}\,{\bf {\bar s}}^2
\,-\,\frac{1}{6}\,\lambda_{4}\,Q^{\prime}
\quad\quad\quad\quad\quad\quad\quad\quad\quad \quad \quad \quad  (3.16)
\end{eqnarray*}
P, Q, $P^{\prime}$ and $Q^{\prime}$ are the four mean-fields,
\begin{eqnarray*}
P\,&=&\,\langle{\bf t}_{m\,i\,\alpha}^{\dag}\,{\bf t}_{m\,i+1\,\alpha}\rangle\,,\quad
Q\,=\,\langle{\bf t}_{m\,i\,\alpha}\,{\bf t}_{m\,i+1\,\alpha}\rangle\,,\\ \\
P^{\prime}\,&=&\,\langle{\bf t}_{r\,i\,\alpha}^{\dag}\,{\bf t}_{l\,i\,\alpha}\rangle\,,\quad
Q^{\prime}\,=\,\langle{\bf t}_{r\,i\,\alpha}\,{\bf t}_{l\,i\,\alpha}\rangle
\end{eqnarray*}
We now perform a Bogolyubov transformation into two new boson operators in 
terms of the {\bf t} operators of the left and right hand ladders as
\begin{eqnarray*}
{\bf \tau}_{1,\,2\, k\, \alpha}\,&=&\,\frac{1}{\sqrt{2}}[ \left( \cosh{\theta_{1,\,2\,k}}{\bf t}_{l\,k\,\alpha}
\,+\,\sinh{\theta_{1,\,2\,k}} \, {\bf t}_{l\,-k\,\alpha}^{\dag}\right) \\ \\
\,& \pm & \, \left (\cosh{\theta_{1,\,2\,k}} \,{\bf t}_{r\,k\,\alpha}
\,+\,\sinh{\theta_{1,\,2\,k}} \,{\bf t}_{r\,-k\,\alpha}^{\dag} \right) ] 
\quad \quad \quad (3.17)
\end{eqnarray*}
These are symmetric (bonding) and antisymmetric (antibonding) combinations of 
the transformations in the left and right ladders. The Hamiltonian (Eq.(3.15)) 
can now be diagonalised to obtain
\begin{eqnarray*}
H_m(\mu,\,{\bf \bar{s}},\,P,\,Q,\,P^{\prime},\,Q^{\prime})\,&=&\,2N\,\left(-\frac{3}{4}\,\alpha\, {\bf \bar{s}}^2\,-\,\mu \,{\bf \bar{s}}^2\,+\,\mu\right)
\,- \, N \, \left(\frac{\alpha}{4} \,- \,\mu\right) \\ \\ \,&-&\,\frac{2}{3}\,\lambda_{2}\,N\,
\left(P^2\,-\,Q^2\right)
\,-\,\frac{1}{6}\,\lambda_{4}\,N\,\left(P^{\prime\,2}\,-\,Q^{\prime\,2}\right)\\ \\
\,&+& \,\sum_{k,\,m=1,\,2} {\bf \omega}_{m\,k} \,\left({\bf \tau}_{m\,k}^{\dag} \,{\bf \tau}_{m\,k}\,+
\,\frac{1}{2}\right)\quad \quad \quad (3.18)
\end{eqnarray*}
where $\omega_{1,\,2\, k}$ is defined as
\begin{eqnarray*}
{\bf \omega}_{1,\,2\,k}\,=\,\sqrt{\left(C\,\mp \,A_k\right)^2\,-\,\left(2\,B_k\,\mp\,D\right)^2 } \quad \quad\quad\quad\quad\quad (3.19)
\end{eqnarray*}
The spin-triplet excitation spectrum of the four-chain ladder consists of two 
branches corresponding to the bonding and antibonding states. The magnitude 
of the splitting of the two branches is determined by $\lambda_3$ and $\lambda_4$. 
Thus the two branches collapse into a single 
branch, when both $\eta\,=\,0,\, \xi\,=\,0$ as in the 
case of a single two-chain ladder.

Eq. (3.19) describes the triplet excitation spectrum in a general parameter 
regime. The parameters ${\bf \mu,\,\bar{s},\,P,\,Q,\,P^{\prime}}$ 
and ${\bf Q^{\prime}\,}$ in the 
excitation spectrum are determined by solving the saddle-point equations:
\begin{eqnarray*}
\langle \,\frac{\delta H_m}{\delta \mu} \,\rangle \, &=& \,0 \,,\quad
\langle\, \frac{\delta H_m}{\delta \bar{s}} \,\rangle \, = \,0 \,,\quad
\langle\, \frac{\delta H_m}{\delta P}\,\rangle\,=\,0\,,\\ \\
\langle\, \frac{\delta H_m}{\delta Q}\,\rangle\,&=&\,0\,,\quad
\langle\, \frac{\delta H_m}{\delta P^{\prime}}\,\rangle\,=\,0\,,\quad
\langle\, \frac{\delta H_m}{\delta Q^{\prime}}\,\rangle\,=\,0
 \quad \quad \quad\quad\quad\quad (3.20)
\end{eqnarray*}
At $T\,=\,0$, the mean-field equations are obtained as
\begin{eqnarray*}
{\bf \bar{s}}^2 \, &=&\, \frac{3}{2}\, + \, \frac{1}{8\,\pi} 
\,\int \, \left(\frac{C\,-\,A_k}{\bf \omega_{1}}\,-\,\frac{C\,+\,A_k}{\bf \omega_{2}}\right)\,dk\\ \\
P\,&=&\,-\,\frac{1}{8\,\pi}\, \int\,\left(\,\frac{C\,-\,A_k}{\bf \omega_{1}}
\,-\,\frac{C\,+\,A_k}{\bf \omega_{2}}\right)\,\cos{k}\,dk\\ \\
Q\,&=&\,-\,\frac{1}{8\,\pi}\,\int\,\left(\frac{2\,B_k\,-\,D}{\bf \omega_{1}}
\,+\,\frac{2\,B_k\,+\,D}{\bf \omega_{2}}\right)\,\cos{k}\,dk\\ \\ 
P^{\prime}\,&=&\,\frac{1}{8\,\pi}\,\int\,\left(\,\frac{C\,-\,A_k}{\bf \omega_{1}}
\,+\,\frac{C\,+\,A_k}{\bf \omega_{2}}\right)\,dk\\ \\
Q^{\prime}\,&=&\,\frac{1}{8\,\pi}\,\int\,\left(\frac{2\,B_k\,-\,D}{\bf \omega_{1}}
\,-\,\frac{2\,B_k\,+\,D}{\bf \omega_{2}}\right)\,dk\\ \\
{\bf \mu}\,&=&\,-\,.75 \,{\bf \alpha}\,+\,\lambda_{1}\,\left(P\,+\,Q\right) \,+\,\frac{\lambda_{3}}{4}\,\left(P^{\prime}\,+\,Q^{\prime}\right)
 \quad \quad \quad \quad(3.21)
\end{eqnarray*}
Fig. 4 shows the spin-triplet excitation spectrum  of the 
four-chain ladder for the exchange interaction strengths 
$\beta\,=\,2\,\gamma\,=\,1$ and $\xi\,=\,2\,\eta\,=\,1\,$, in units of $\alpha$. 
Fig. 5 shows the spin-gap of the four-chain ladder versus $\xi$ 
for $\beta\,=\,1$, $\eta\,=\,\frac{\xi}{2}$ and 
$ \gamma\,=\,0.5$, in units of \(\alpha\).

The ground state energy of the ladder system in the general
parameter regime is given by 

\begin{eqnarray*}
E_g\,&=&\,2N\,\left(-\frac{3}{4}\,\alpha\, {\bf \bar{s}}^2\,-\,\mu \,{\bf \bar{s}}^2\,+\,\mu\right)
\,- \, N \, \left(\frac{\alpha}{4} \,- \,\mu\right) \,-\,\frac{2}{3}\,\lambda_{2}\,N\,
\left(P^2\,-\,Q^2\right)\\ \\
\,&-&\,\frac{1}{6}\,\lambda_{4}\,N\,\left(P^{\prime\,2}\,-\,Q^{\prime\,2}\right)
\,+\, \frac{1}{2}\,\sum_{k,\,m=1,\,2} {\bf \omega}_{m\,k} 
\end{eqnarray*}
The parameter regime includes the point ${\bf \alpha\,\geq\,3\,\beta,\,
\beta\,=\,\gamma\,=\,\xi\,=\,\eta}$ at which the ground state and
the corresponding energy are exactly known. For these parameter
values, ${\bf \lambda_1\,=\,0,\,\lambda_3\,=\,0}$ and ${\bf \lambda_2\,=\,2\,\beta,\,
\lambda_4\,=\,2\,\beta}$. Also, the four mean-fields {\bf P, Q,
}${\bf P^\prime, Q^\prime}$ are zero, ${\bf
\bar{s}^2\,=\,1}$ and ${\bf \mu\,=\,-.75\,\alpha}$.

From Eq. (3.19), one then obtains a single excitation spectrum of
energy ${\bf \omega_k\,=\,\alpha}$, i.e., the spectrum is
dispersionless. The ground state energy $E_g$ becomes
\begin{eqnarray*}
E_g\,&=&\,2\,N\,\left(\,-\frac{3}{4}\,\alpha\,\bar{s}^2\,-\,\mu\,\bar{s}^2\,+\,\mu\,\right)
\,-\,N\,\left(\frac{\alpha}{4}\,-\,\mu\,\right)\,+\,\frac{1}{2}\,
\sum_{k,\,m\,=\,1,\,2}\,{\bf \omega_{m\,k}}\\ \\
\,&=&\,-\,\frac{3}{2}\,{\bf \alpha}\,N
\end{eqnarray*}
which is equal to the exact ground state energy. The mean-field
theory based on the bond operator formalism thus reproduces the
correct ground state energy in the appropriate limit of the
coupling parameters.

Next we consider a periodic array of ladders, i.e., consider the full square 
lattice with exchange interactions as specified before. The problem of 
interest is to find the value of the interladder interaction strength \({\bf \xi}\) 
at which the spin-gap disappears. For the usual square lattice $S \, = \, 1/2$ HAFM with only n.n. 
interactions, long-range AFM order exists in the ground state and the spin-gap is 
expected to vanish at a critical value of the interladder exchange interaction 
$\xi \, (\eta\,= \,0,\, \gamma \, = \,0 \,$ in this case ).
The value obtained by Gopalan et al is $\xi$ = 0.25. The spin-ladder model constructed 
by us has not only n.n. but n.n.n. (along the rungs) as well as diagonal 
interactions. It is of interest to determine whether for this model also the 
spin-gap vanishes at a critical value of $\xi$. 
Using the formalism already developed, we obtain the self-consistent equations 
\begin{eqnarray*}
{\bf {\bar s}}^2\,&=&\,\frac{3}{2}\,+\,\frac{1}{8\,\pi^2}\,\int\,\int\,\frac{C_k\,-\,A_k}{\bf \omega}\,d{\bf k}\\ \\
P\,&=&\,-\,\frac{1}{8\,\pi^2}\,\int\,\int\,\frac{C_k\,-\,A_k}{\bf \omega}\,\cos{k_x}\,d{\bf k}\\ \\
Q\,&=&\,-\,\frac{1}{8\,\pi^2}\,\int\,\int\,\frac{2\,B_k\,-\,D_k}{\bf \omega}\,\cos{k_x}\,d{\bf k}\\ \\
P^{\prime}\,&=&\,\frac{1}{8\,\pi^2}\,\int\,\int\,\frac{C_k\,-\,A_k}{\bf \omega}\,\cos{k_y}\,d{\bf k}\\ \\
Q^{\prime}\,&=&\,\,\frac{1}{8\,\pi^2}\,\int\,\int\,\frac{2\,B_k\,-\,D_k}{\bf \omega}\,\cos{k_y}\,d{\bf k}\\ \\
{\bf \mu}\,&=&\,-\,.75 \,{\bf \alpha}\,+\,\lambda_{1}\,\left(P\,+\,Q\right) \,+\,\frac{\lambda_{3}}{2}\,\left(P^{\prime}\,+\,Q^{\prime}\right)\quad\quad\quad(3.22)
\end{eqnarray*}where $ A_k,\,B_k,\,C_k,\,$ and $D_k$ are
\begin{eqnarray*}
A_k\,&=&\,\frac{\alpha}{4}\,-\,{\bf \mu}\,+\,\left( \lambda_{1}\,{\bf {\bar s}}^2\,
+\,\frac{2}{3}\,\lambda_{2}\,P\right)\,\cos{k_x}\,,\\ \\
B_k\,&=&\,\left(\frac{1}{2}\,\lambda_{1}\,{\bf {\bar s}}^2\,-\,\frac{1}{3}\,\lambda_{2}\,Q\right)\,\cos{k_x}\,,\\ \\
C_k\,&=&\,\left(\frac{1}{2}\,\lambda_{3}\,{\bf {\bar s}}^2\,+\,\frac{1}{3}\,\lambda_{4}\,P^{\prime}\right)\,\cos{k_y}\,,\\ \\
D_k\,&=&\,\left(\frac{1}{2}\,\lambda_{3}\,{\bf {\bar s}}^2\,-\,\frac{1}{3}\,\lambda_{4}\,Q^{\prime}\right)\,\cos{k_y}
\quad\quad\quad\quad\quad\quad\quad\quad(3.23)
\end{eqnarray*}
Also, ${\bf k}$ is a two-dimensional wave vector with components $k_x$ (along 
the ladder axis) and $k_y$ (across the ladders). The excitation spectrum 
${\bf \omega}_{\bf k}$ is given by
\begin{eqnarray*}
{\bf \omega}_{\bf k} \, = \,\sqrt{\left( C_k\,-\,A_k\right)^2\,-\,\left(2\,B_k\,-\,D_k\right)^2} \quad \quad \quad \quad\quad(3.24)
\end{eqnarray*}
The excitation spectrum ${\bf \omega_k}$ has a minimum at ${\bf k} \, = \, (\pi,\,0)$ .
Fig. 6 shows a plot of the spin-gap $ {\Delta} $ versus $ \xi \, $
for ${\bf \beta \,=\,1,\,\gamma\,=\,0.5\,}$ and ${\bf \eta\,=\,\frac{\xi}{2}}$, in units of ${\bf \alpha}$.
The spin-gap $\Delta$ vanishes for ${\bf \xi\,=\,0.33}$. 
\section*{IV. Conclusions}
We have constructed spin ladders with odd and even 
number of chains for which in a particular parameter regime the exact 
ground state can be written down. It can further be shown rigorously that 
the excitation spectrum is gapless (with a gap) for odd (even) number of 
chains. 
The mean-field theory based on the bond operator formalism has been  
applied to ladders with an even number of chains in a general parameter 
regime. Both the formalism and the results obtained are similar to those of 
Gopalan et al\cite{Gopalan} for spin ladders which differ from ours in that the 
diagonal and n.n.n. interactions along the rung are absent. 
One significant difference is in the inclusion of terms containing four 
triplet operators in our mean-field theory. For the ladder models considered by 
Gopalan et al these terms have a negligible contribution and so have been 
ignored. In the present case, the terms can no longer be neglected. 
The results of Gopalan et al \cite{Gopalan} can be recovered from 
our results by putting ${\bf \gamma \, = \, 0, \,\eta \, = \, 0}$. 
Inclusion of these extra interactions has the effect of 
renormalising the original coupling parameters of the Hamiltonian when expressed in 
terms of the singlet and triplet operators. For the four-chain spin ladder considered 
in Ref.4, the coupling parameters ${\bf \beta \, - \, \gamma}$ and ${\bf \eta \, - \, \xi}$ in 
$Eq.\left(3.14\right)$ are ${\bf \beta}$ and - ${\bf \xi}$ respectively. 
The inclusion of frustrating further-neighbour 
interactions in our model has the effect of increasing the spin-gap. 
For the periodic ladder, the spin-gap vanishes at ${\bf \xi}$ = 0.33 ( Fig. 6 ). 
The decrease of the spin-gap with ${\bf \xi}$ is explained by the 
delocalisation of the singlets across the ladders. The decrease of the 
gap is faster than that of a four-chain ladder.

The mean-field theory based on the bond operator formalism reproduces the 
exact ground state energy in the appropriate limit. The same is true for two 
other spin models in 1d and 2d, namely, the Majumdar-Ghosh 
chain\cite{majum} and the $J_1-J_2-J_3-J_4-J_5$ model 
proposed by Bose and Mitra\cite{bose_mitra,bhaumik_bose}. 

The ground state of both the 
models can be determined exactly at particular values of the parameters. The 
ground states consist of a periodic arrangement of dimers. Mean-field 
theory based on the bond operator formalism determines the ground state 
energy correctly in the exactly-solvable limit. In the same limit, the 
mean-field theory yields a dispersionless excitation spectrum for both 
the spin models. This is also true for the ladder spin system signifying 
that the three spin models share common features.

The sawtooth chain which describes the exchange interactions along the rungs 
of the ladder system has been studied earlier by Kubo\cite{Kubo}. The ground 
state of the chain is doubly degenerate and the spin dynamics is described 
in terms of kink, antikink excitations.

Consider the parameter regime in which the exact ground state of the ladder 
system is known. The sawtooth chain interactions are now (Fig. 2) $\eta = \xi 
= \frac{\alpha}{3}$. In this case the ground state is nondegenerate with spin 
singlets forming along the stronger bonds. Kink, antikink excitations which 
can be considered as spin defects separating the two degenerate ground states 
are absent in this case. Spin excitations are now created if one of the 
singlets is replaced by a triplet and the triplet is allowed to propagate. 
In sawtooth chain with doubly degenerate ground states these excitations have 
a higher energy than the kink, antikink excitations.
\vskip 2cm
{\bf Acknowledgement}
\newline
We thank Sujit Sarkar for computational help.
\newpage
\section*{Figure Captions}
\begin{description}
\item[Fig. 1]  
    The two-chain ladder with rung, horizontal and diagonal exchange 
interactions of strength ${\bf \alpha,\,\beta }$ and ${\bf \gamma}$ and 
depicted by dashed ( bold ), solid ( bold ), and solid ( thin ) lines 
respectively .
\item [Fig. 2]
    The sawtooth chain with three different interactons of strength 
${\bf \alpha ,\, \xi}$ and ${\bf \eta}$ and depicted by dashed ( bold ), 
dashed ( thin ) and dot-dashed lines 
respectively.
\item [Fig. 3(a)]
    The three-chain ladder with five different interactions of strength 
${\bf \alpha,\,\beta,\,\gamma,\,\xi}$ and ${\bf \eta}$ and depicted by 
dashed ( bold ), solid ( bold ), solid ( thin ), dashed ( thin ) and 
dot-dashed lines respectively.
\item [Fig. 3(b)]
    The four-chain ladder with five different interactions of strength 
${\bf \alpha,\,\beta,\,\gamma,\,\xi}$ and ${\bf \eta}$.
\item [Fig. 4]
   The triplet excitation spectrum [ bonding and antibonding states of Eq. (3.19) ] 
of the four-chain ladder with exchange interaction 
strengths ${\bf \,\beta\,=\,2\,\gamma\,=\,1}$ and ${\bf \xi\,=\,2\,\eta\,=\,1}$, 
in units of ${\bf \alpha}$.
\item [Fig. 5]
   The spin-gap ${\bf \Delta\,}$ of the four-chain ladder versus $\xi$,  
for ${\bf \eta\,=\,\frac{\xi}{2},\,\beta\,=\,1}$ and ${\bf \gamma\,=\,0.5}$, 
in units of ${\bf \alpha}$.
\item [Fig. 6]
   The spin-gap ${\bf \,\Delta\,}$ of the peroidic ladder versus $\xi$,  
for ${\bf \eta\,=\,\frac{\xi}{2},\,\beta\,=\,1}$ and ${\bf \gamma\,=\,0.5}$,
 in units of ${\bf \alpha}$ . The spin-gap vanishes for ${\bf \xi\,=\,0.33}$.
\end{description}

\end{document}